\documentclass{article}

\usepackage{arxiv}

\usepackage[utf8]{inputenc} 
\usepackage[T1]{fontenc}    
\usepackage[hidelinks]{hyperref}       
\usepackage{url}            
\usepackage{booktabs}       
\usepackage{amsfonts}       
\usepackage{amsmath}
\usepackage{amssymb}
\usepackage{nicefrac}       
\usepackage{microtype}      
\usepackage{lipsum}
\usepackage{graphicx}
\graphicspath{ {./images/} }
\usepackage{subfig}
\usepackage{cite}
\usepackage{xcolor}
\definecolor{mBlue}{RGB}{31, 119, 180}
\definecolor{mOrange}{RGB}{255, 127, 14}
\definecolor{mGreen}{RGB}{44, 160, 44}
\definecolor{mRed}{RGB}{214, 39, 40}
\newcommand\solidXXLrule[1][1cm]{\rule[0.001ex]{#1}{4pt}}

\title{Estimation of motion blur kernel parameters using regression convolutional neural networks$^{1,2}$}

\author{
  Luis G.~Varela, Laura E.~Boucheron, Steven Sandoval, David Voelz, and Abu Bucker Siddik \\
  Klipsch School of Electrical and Computer Engineering\\
  New Mexico State University\\
  Las Cruces, NM 88001, USA \\
  \texttt{\{varelal,lboucher,spsandov,davvoelz,siddik\}@nmsu.edu} \\
}

\begin{document}
\maketitle
\begin{abstract}
Many deblurring and blur kernel estimation methods use a maximum a posteriori (MAP) approach or deep learning-based classification techniques to sharpen an image and/or predict the blur kernel. We propose a regression approach using convolutional neural networks (CNNs) to predict parameters of linear motion blur kernels, the length and orientation of the blur. We analyze the relationship between length and angle of linear motion blur that can be represented as digital filter kernels.  A large dataset of blurred images is generated using a suite of blur kernels and used to train a regression CNN for prediction of length and angle of the motion blur.  The coefficients of determination for estimation of length and angle are found to be greater than or equal to 0.89, even under the presence of significant additive Gaussian noise, up to a variance of 10\% (SNR of 10 dB).  Using our estimated kernel in a non-blind image deblurring method, the sum of squared differences error ratio demonstrates higher cumulative histogram values than comparison methods, with most test images yielding an error ratio of less than or equal to 1.25.
\end{abstract}

\footnotetext[1]{The official version of this paper appears as Luis G. Varela, Laura E. Boucheron, Steven Sandoval, David Voelz, Abu Bucker Siddik, ``Estimation of motion blur kernel parameters using regression convolutional neural networks,'' J. Electron. Imaging 33(2), 023062 (2024); doi: 10.1117/1.JEI.33.2.023062.}
\footnotetext[2]{Copyright 2024 Society of Photo-Optical Instrumentation Engineers. One print or electronic copy may be made for personal use only. Systematic reproduction and distribution, duplication of any material in this paper for a fee or for commercial purposes, or modification of the content of the paper are prohibited.}

\section{Introduction}
Linear motion blur has been studied as a model for camera shake, camera platform movement, and moving objects during imaging~\cite{Camera_shake,motion_blur}. The studies presented within were designed as an initial exploration of a regression convolutional neural network (CNN) to estimate linear motion blur parameters from blurry images. The effects of atmospheric turbulence in an image are generally represented by a spatially-varying blur~\cite{atmospheric,thiebaut2016spatially}, while recent work has demonstrated that spatially-varying blurred images can be modeled as a superposition of locally linear motion blurs~\cite{bahat}. One way to parametrize linear motion blur kernels is by length and orientation. In this work, we study the ability to accurately estimate the length and angle parameters of a uniform linear motion blurred image. The results may then be used as a foundation upon which to build methods to estimate spatially-varying blur parameters and to serve as a baseline in subsequent studies.

A uniform blurred image can be described by 
\begin{equation}
    \label{eq:blurry}
    I_{\text{b}} = K * I_{\text{s}} + N,
\end{equation}
where $I_{\text{b}}$ is the blurred image, $K$ is the blur kernel, $I_{\text{s}}$ is the sharp latent image, $N$ is additive noise, and $*$ is the convolution operator. This formulation assumes a uniform blur because the same kernel $K$ is applied across the entire image.

The main contributions of this paper are a thorough exploration of linear motion blur kernels and development of a regression CNN for blind prediction of motion blur parameters from blurry images.  We study the space of motion blur parametrized by length and angle, specifically the subset of linear motion blur that can be described as 2D discrete motion blur kernels.  Furthermore, we include motion blur kernels that are not described as square odd-sized kernels and study complications that arise in defining 2D discrete motion blur kernels. To that end we create a dataset of blurred images spanning a large range of possible blur kernels for use in training and testing a regression CNN based on the VGG16~\cite{simonyan2014very} network to predict linear motion blur kernel parameters. The network is analyzed using the coefficient of determination $R^2$ score, which is a common metric to quantify performance of regression analysis, as well as a deconvolved error ratio (the ratio of the error when deconvolving using the predicted blur kernel to the error when deconvolving using the actual blur kernel)~\cite{levin2009understanding}.

The organization of this paper is as follows. Section~\ref{sec:RW} gives a discussion on previous work including classical deblurring and deep learning deblurring methods. Section~\ref{sec:Blur-Dataset} contains a detailed description of the exploration of the linear blur kernel parameter space as well as the creation of a blurred dataset for deep learning training. Section~\ref{sec:blur_prediction} presents the proposed regression CNN for blur parameter prediction and Sec.~\ref{sec: Experiment and results} presents results of our regression prediction method along with comparison to previous work.  Finally, in Sec.~\ref{sec:conclusions} we provide a conclusion and briefly discuss our future work.

\section{Related Work}
\label{sec:RW}
Deblurring and kernel blur estimation have been extensively studied in computer vision with applications of recovering the sharp image from blurry images caused by camera shake, fast moving objects in frame, or atmospheric turbulence. Prior to deep learning methods, many researchers used a maximum a posteriori (MAP) approach for both blur kernel estimation and deblurring. Two commonly used varieties of MAP are implicit regularization as seen in~\cite{Xu_2013_CVPR,BD_Krishnan} and explicit edge prediction based approaches like those in~\cite{FMD,Radon_T,MONEY2008302,jia_2007_ieee,Camera_shake}. While many approaches have used MAP to estimate both the latent image and the blur kernel,  Levin et al.~\cite{Levin_2011_ieee, levin2009understanding} prove that this approach tends to favor the no-blur explanation (i.e., that the kernel is an impulse and the ``deblurred'' image is the blurry image) instead of converging to the true blur kernel and latent sharp image. Moreover, it is advocated in~\cite{Levin_2011_ieee,levin2009understanding} that estimating only the blur kernel is a better approach since there are fewer parameters to determine than if one were to also estimate the latent sharp image.  
Levin et al.~\cite{Levin_2011_ieee,levin2009understanding} use an expectation-maximization (EM) framework to optimize kernel prediction with either a Gaussian or sparse prior. More recently methods have considered deep learning approaches to estimate the blur kernel~\cite{yan_2016_ieee}, estimate the latent sharp image~\cite{zhang_2018_ieee}, or both~\cite{Li_2018_CVPR}.

In the first variety of MAP-based methods, edge-based approaches in a MAP framework use extracted edges as an image prior in the MAP optimization. Cho et al.~\cite{FMD} introduced a gradient computation using derivatives of the image to compute edges. Money \& Kang~\cite{MONEY2008302} used shock filters for edge detection. Cho et al.~\cite{Radon_T} used the Radon transform for edge based analysis; image deblurring may use a MAP algorithm or the inverse Radon transform informed by the detected edges. Jia~\cite{jia_2007_ieee} used object boundary transparency as an estimation of edge location. Fergus et al.~\cite{Camera_shake} introduced natural image statistics defined by distributions of image gradients as a prior.

In the second variety of MAP-based methods, implicit regularization MAP approaches use different regularization terms to enforce desired image priors. Xu et al.~\cite{Xu_2013_CVPR} incorporated a regularization term to approximate the $L_{0}$ cost which improves computational speeds over alternative implicit sparse regularizations. The framework in~\cite{Xu_2013_CVPR} alternated between estimating the latent sharp image and the blur kernel in each iteration. Krishnan et al.~\cite{BD_Krishnan} used a ratio of $L_1 / L_2$ norms as a regularization to estimate the kernel. This helps with the attenuation of high frequencies that blur introduces in an image. 

Although many of the previous works discussed above involve the estimation of generalized blur kernels, some work specializes in motion blur kernel prediction. Whyte et al.~\cite{whyte2012non} proposed a new method to estimate motion blur by parametrizing a geometric model in terms of rotational velocity of the camera during exposure time.  They modified the Fergus et al.~\cite{Camera_shake} 
algorithm to implement non-uniform deblurring. Hirsch et al.~\cite{hirsch_2011_ieee} combined projective motion path blur models with efficient filter flow~\cite{hirsch2010efficient} to estimate motion blur at a patch level and deblur by modifying the Krishnan \& Fergus~\cite{krishnan2009fast} algorithm.

Recent methods using deep learning have used various architectures including convolutional neural networks (CNNs)~\cite{Sun_2015_CVPR,Xu_2018_ieee,Li_2018_CVPR,yan_2016_ieee,nasonov2022}, encoder-decoder networks~\cite{carbajal2021}, generative adversarial networks (GANs)~\cite{zhang_2018_ieee}, and fully convolutional networks (FCNs)~\cite{Gong_2017_CVPR} to tackle the problems of blur kernel prediction and deblurring. Sun et al.~\cite{Sun_2015_CVPR} used a CNN to classify the best-fit motion blur kernel for each image patch. They used 73 different motion blur kernels and were able to expand to 361 kernels by rotation of their input image. Xu et al.~\cite{Xu_2018_ieee} used a CNN to recover sharp edges from blurred images and then estimated the blur kernel from the blurry and recovered edges via a MAP optimization.  Li et al.~\cite{Li_2018_CVPR} used a mixture of deep learning and MAP estimation to deblur an image, using a binary classification CNN trained to predict the probability of whether the input is blurry or sharp. The classifier is used as a regularization term of the latent image prior in a MAP framework to deblur. Yan \& Shao~\cite{yan_2016_ieee} used a two step process to predict blur parameters: first, a deep neural network classified the type of blur and second, a general regression neural network predicted the parameters of the blur. Nasonov and Nasonova~\cite{nasonov2022} used a CNN to predict length and angle for linear motion blur with several formulations of the output values, although they noted difficulty in the simultaneous estimation of length and angle.  Carbajal et al.~\cite{carbajal2021} used an encoder and two decoders to predict motion kernel bases for the image and mixing coefficients for each pixel.  Zhang et al.~\cite{zhang_2018_ieee} used a conditional GAN to deblur images, combining an adversarial loss and a perceptually-motivated content loss.  Gong et al.~\cite{Gong_2017_CVPR} used an FCN to predict a motion flow map where each pixel has its own motion vector estimate.  While many of these approaches are able to handle spatially varying blur, we focus on a thorough exploration of motion kernel parameters in uniform blur; this will serve as a framework for spatially varying blur in future work. 

There are three main contributions to this work.  First,
we study complications that arise in defining a motion blur kernel and, in particular, we analyze which motion blur kernels exist within different kernel shapes rather than using square odd-sized kernels as often implicitly assumed in the implementation---likely for ease of coding. Second, by exploring all motion blur kernel possibilities for a suite of length and angle combinations, we train a CNN which uses regression prediction instead of classification. By formulating the prediction as a regression, we train two output nodes (one for length and one for angle) which can span the full range of possible parameter values and can be trained on any granularity of parameters.  Use of a classification network would necessitate 13,034 output nodes for the granularity considered here (99 lengths and 180 angles) and modification of the network architecture to train on any other granularity or post-hoc computation (e.g., as in~\cite{Sun_2015_CVPR}) to define predictions at another granularity.  Use of regression also may alleviate some of the issues in simultaneous prediction of length and angle noted in~\cite{nasonov2022}. Third, we expand the range of additive noise beyond that studied for other methods~\cite{Levin_2011_ieee,Radon_T,Xu_2018_ieee,Li_2018_CVPR,BD_Krishnan,whyte2012non} to analyze Gaussian noise up to a variance of 10\% (10 dB SNR).

\section{Blur Kernels and Blurred Dataset} 
\label{sec:Blur-Dataset}
A linear blur kernel can be parametrized by the length and orientation angle of a line. To formulate a discrete blur kernel for application to a digital image using Eq.~(\ref{eq:blurry}), we specifically distinguish between a continuous line and a discrete pixel line. Next, we explore angle and length combinations that exist within linear motion blur kernels. Finally, we describe how the 2014 COCO dataset~\cite{lin2014microsoft} is used to create a new blurred dataset for deep learning training and testing.

\subsection{Continuous and Pixel Lines}
\begin{figure}[t]
    \centering
    \includegraphics[width=.45\linewidth]{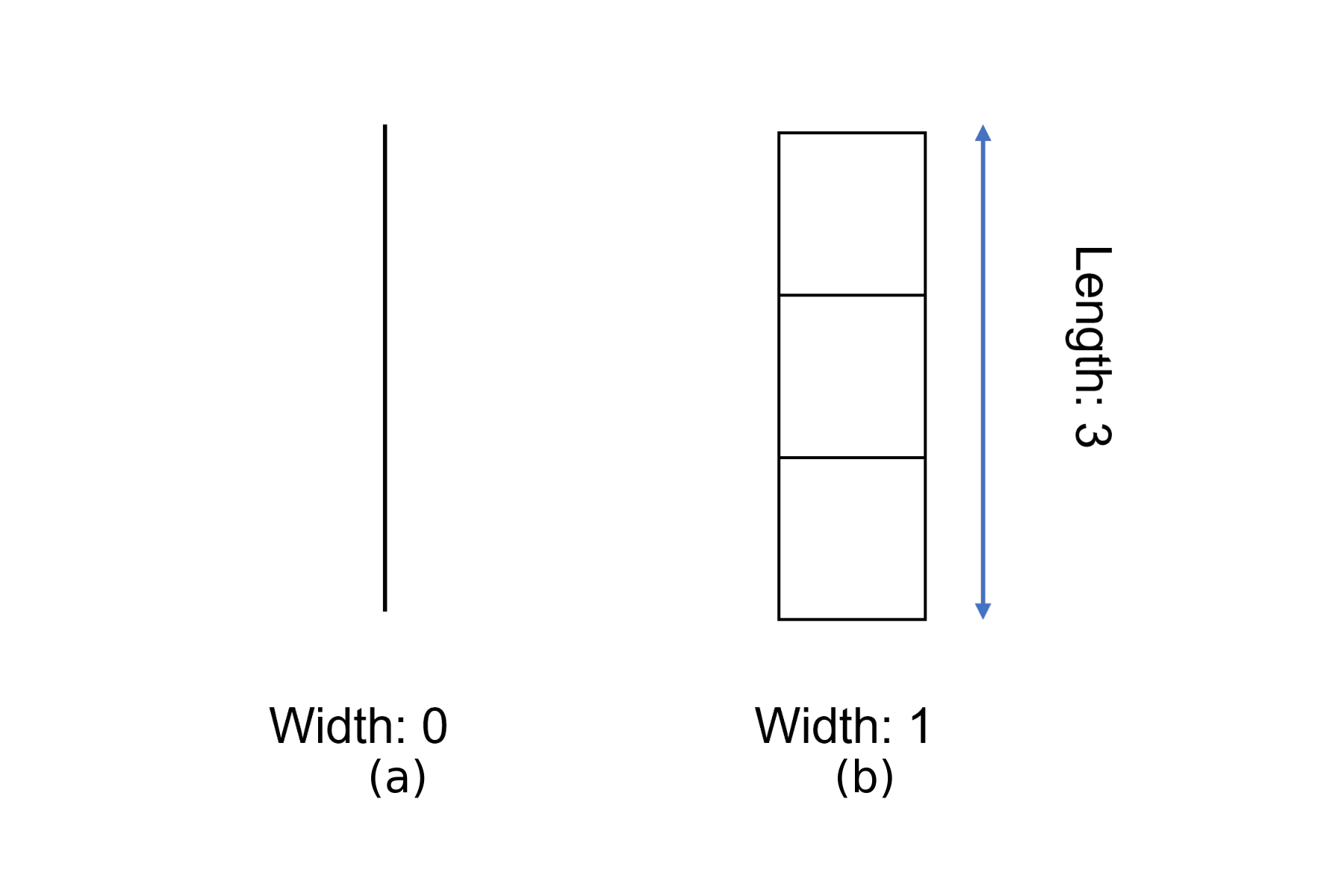}
    \caption{(a) A continuous line with parameters $(r,\theta)=(3,90^\circ)$ and (b) pixel line with parameters $(r,\phi)=(3,90^\circ)$.  Note that the pixel line has a width of 1 due to its definition on a 2D discrete grid.}
    \label{fig:line}
\end{figure}

Here, the term ``line'' refers to a line segment, $r$ denotes a Euclidean distance (or a continuous or discrete line), $r_\infty$ is a Chebyshev distance for a discrete line, $\theta\in(-90,90]$ denotes the orientation (angle) of a continuous line with respect to the horizontal axis in units of degrees, and $\phi\in(-90,90]$ is the angle of a discrete pixel line with the same conventions.  
For a given length $r$, the set of possible continuous line angles is denoted $\Theta\in(-90,90]$ and the set of possible discrete pixel line angles is $\Phi\subseteq\Theta$.  A line in a 2D continuous domain $\mathbb{R}^2$ is a 1D object with zero width and length $r$.
On the other hand, a pixel line in a 2D discrete domain $\mathbb{Z}^2$ is defined on a discrete pixel grid which gives the line a non-zero width and length $r$. Figure~\ref{fig:line} shows a continuous line with parameters $(r,\theta)=(3,90)$ and a pixel line with parameters $(r,\phi)=(3,90)$.

\begin{figure}[t]
    \centering
    \includegraphics[]{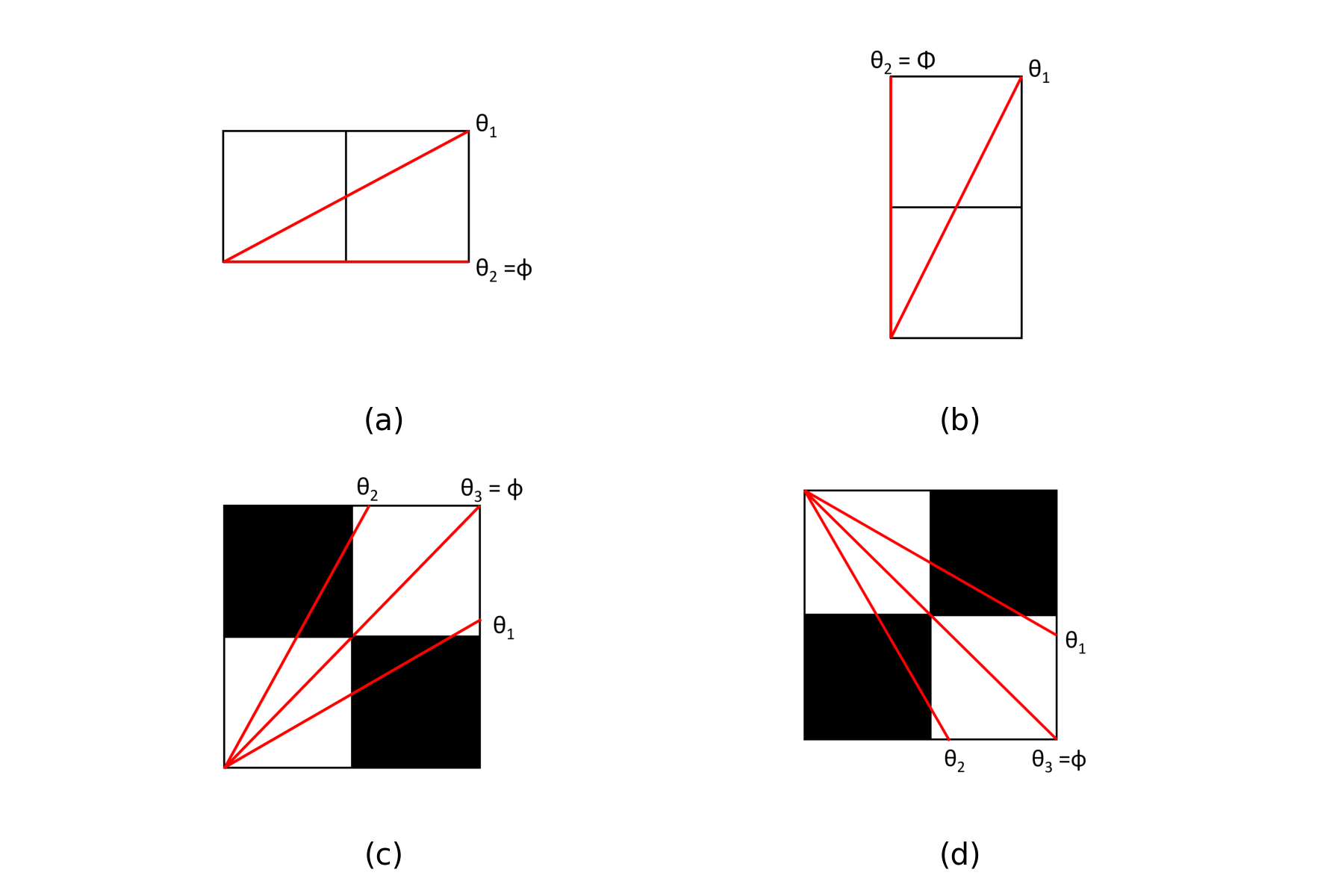}
    \caption{Multiple continuous lines can be represented by the same pixel line of Chebyshev length $r_\infty=2$.  Example continuous lines are shown in red, overlaid on the pixel line kernel (the squares) that describes those continuous lines, where white squares (pixels) correspond to the line and black squares (pixels) correspond to the absence of a line.  (a) Two continuous lines (with $\theta_1=30^\circ$ and $\theta_2=\phi=0^\circ$) described by the same $\phi=0^{\circ}$ pixel line. (b) Two continuous lines (with $\theta_1=60^\circ$ and $\theta_2=\phi=90^\circ$) described by the same $\phi=90^\circ$ pixel line. (c) Three continuous lines (with $\theta_1=31^\circ$, $\theta_3=\phi=45^\circ$, and $\theta_2=59^\circ$ described by the same $\phi=45^{\circ}$ pixel line. (d) Three continuous lines (with $\theta_1=-31^\circ$, $\theta_3=\phi=-45^\circ$, and $\theta_2=-59^\circ$ described by the same $\phi=-45^{\circ}$ pixel line.}
    \label{fig:pix-lines}
\end{figure}

One limitation in defining pixel lines is the interrelation between length and angle. In essence, a shorter pixel line will have a limited number of unambiguous angles. Figure~\ref{fig:pix-lines}(a) illustrates how multiple continuous lines with angle $\theta\in[0,30]$ can be interpreted by the same horizontal pixel line of length $r_\infty=2$. Similarly, a pixel line of length $r_\infty=2$ and angle $\phi=45$ [Fig.~\ref{fig:pix-lines}(c)] can describe many intermediate lines with $\theta\in(30,60)$.  The four pixel lines in Fig.~\ref{fig:pix-lines} show the only pixel lines that can be described for length $r_\infty=2$.  This means that, for shorter length motion blurs, there will be gaps in the angles that can be represented.  For example, a length $r_\infty=2$ pixel line can only have angles $\phi\in\{-45,0,45,90\}$ as shown in Fig.~\ref{fig:pix-lines}.   

Conventional convolutional kernels are typically assumed to be square and odd in dimension.  The preference for square kernels is related to the horizontal and vertical symmetry of features. The preference for odd-sized kernels is to have an unambiguous center point, as the center point is commonly assumed to be associated with the pixel being processed. 
The assumption of square or odd-sized kernels in the blur model of Eq.~(\ref{eq:blurry}) is not mathematically necessary, however, and we consider blur kernels that may be non-square and/or even in one or more dimension.  

\subsection{Blur Kernel Creation}
\label{sec:blur_kernel}

\begin{figure}[t]
\centering
    \includegraphics[width=\textwidth]{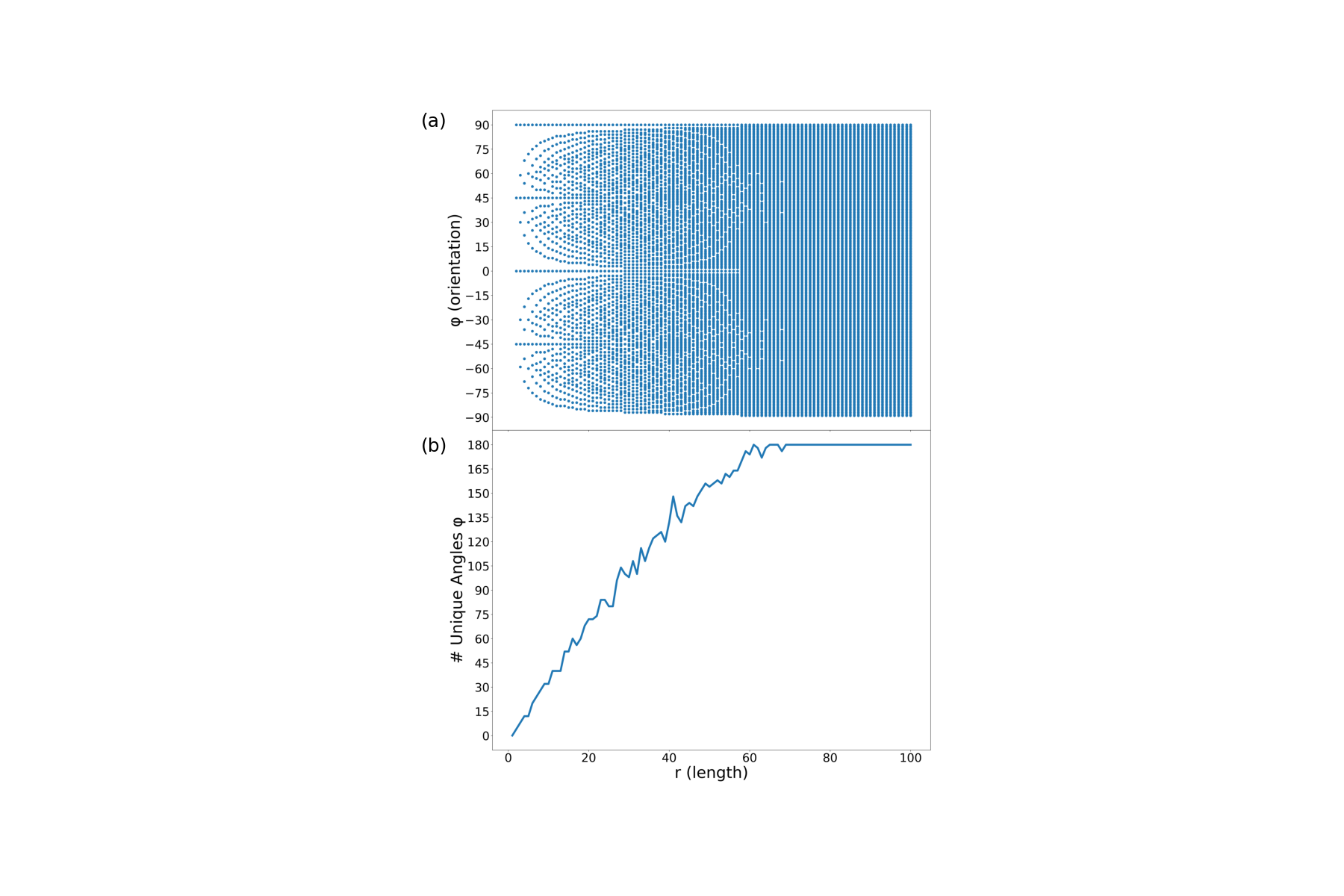}
    \caption{(a) Unique length and angle combinations $(r,\phi)_\text{u}$ for all possible discrete blur kernels for $r\in[2,100]$ and $\phi\in(-90,90]$.  Shorter length kernels have many fewer possible angles that they can represent, including the four possible angles of $\phi\in\{-45, 0, 45, 90\}$ for $r=2$. (b) Number of angles that can be represented for a given kernel length.  Shorter length kernels have many fewer possible angles that they can represent.}
    \label{fig:pix-range}
\end{figure}

To explore the gaps in angles resulting from representation of continuous lines as pixel lines, we assume a range of Euclidean lengths $r=2,3,\ldots,100$ and angles $\theta=-89,-88,\ldots,90$, and gather the resulting unique $(r,\phi)$ discrete line parameter pairs, illustrated in Fig.~\ref{fig:pix-range}
.  We assume the $h\times w$ kernel corresponding to an $(r,\theta)$ continuous line has $h$ and $w$ defined as
\begin{equation}
    h = \begin{cases}
    \lceil r\cos(\theta)\rceil, & \cos(\theta) \neq 0 \\
    1, & \cos(\theta) = 0 \\
    \end{cases},
\end{equation}
\begin{equation}
    w = \begin{cases}
    \lceil r\sin(\theta)\rceil, & \sin(\theta) \neq 0 \\
    1, & \sin(\theta) = 0 \\
    \end{cases},
\end{equation}
where $\lceil \cdot \rceil$ is the ceiling operator. The conditions $\cos(\theta)=0$ and $\sin(\theta)=0$ explicitly define a height or width of one for the horizontal and vertical cases, respectively. We use $r$ as the Euclidean distance labels and draw an $(r,\theta)$ continuous line (e.g., the red lines in Fig.~\ref{fig:pix-lines}) through an $h\times w$ 2D discrete pixel grid (e.g., the $1\times2$, $2\times1$, and $2\times2$ pixel grids in Fig.~\ref{fig:pix-lines}), resulting in a discrete pixel line (e.g., the white pixels in Fig.~\ref{fig:pix-lines}). The ceiling operator introduces a quantization error; since $r$ may result in a line ending mid-pixel, the ceiling operator corresponds to the choice to use the pixel where $r$ ends as part of the motion blur kernel. The worst case for the quantization error is for angles $\theta=\pm45$ with the Euclidean and Chebyshev distances differing by a factor of $\sqrt{2}$. 

We use the \verb+line+ function from the \verb+skimage.draw+ library in Python to draw a continuous line through the pixel grid, resulting in a discrete pixel line consistent with the Bresenham digital line algorithm~\cite{bresenham1998algorithm}. If $\theta$ is positive, the line is drawn from the lower-left corner $(h,0)$ to the upper-right corner $(0,w)$, e.g., Fig.~\ref{fig:pix-lines}(c), and if $\theta$ is negative, the line is drawn from the upper-left corner $(0,0)$ to the lower-right corner $(h,w)$, e.g., Fig.~\ref{fig:pix-lines}(d).

For each length $r=2,3,\ldots,100$, we generate a discrete blur kernel associated with a continuous line of length $r$ and angle $\theta=0,1,\ldots,90$.  Noting that multiple angles $\theta$ may result in the same discrete blur kernel (see Fig.~\ref{fig:pix-lines}), a single discrete angle $\phi$ must be defined for each unique discrete blur kernel for use as a training label.  We thus quantize the continuous angle $\theta$ for a given blur kernel as 
\begin{equation}
    \phi = \begin{cases}
    0, & 0\in\Theta \\
    90, & 90\in\Theta \\
    \lceil \text{median}(\Theta) \rceil, & \text{else}
    \end{cases},
    \label{eq:phi}
\end{equation}
where $\Theta$ is the set of continuous angles $\theta$ resulting in the given blur kernel and where $\lceil\cdot\rceil$ is the ceiling operator necessary to define an integer angle in cases that $\Theta$ contains an even number of elements.  Note that this definition of $\phi$ has two special cases. If $0\in\Theta$, the kernel is consistent with a horizontal line [see Fig.~\ref{fig:pix-lines}(a)] and we assign $\phi=0$. Similarly, if $90\in\Theta$, the kernel is consistent with a vertical line [see Fig.~\ref{fig:pix-lines}(b)] and we assign $\phi=90$.  These special cases avoid the median operator assigning an erroneous label to a horizontal or vertical line. 

We compute all possible $(r,\phi)$ combinations by generating the blur kernel for $r=2,3,\ldots,100$ and $\theta=0,1,\ldots,90$ and computing the resulting blur kernel angle $\phi$ using Eq.~(\ref{eq:phi}).  We span only positive angles $\theta\in[0,90]$ since kernels with negative angles will be symmetric versions of the kernels with positive angles. The resulting values $r$ and $\phi$ that yield unique discrete pixel lines are denoted as $(r,\phi)_\text{u}$.  These parameter values are also used as labels to train a network to predict the blur parameters from a blurry image (Sec.~\ref{sec:blur_prediction}) using $(r,\phi)=(1,0)$ as labels for a non-blurred image.

Figure~\ref{fig:pix-range} 
illustrates the $(r,\phi)_\text{u}$ combinations as a scatter plot, where we note the existence of gaps where a continuous line cannot be described by a pixel line. As expected, gaps are larger for smaller lengths, indicating limited unique pixel lines (blur kernels) for shorter blur lengths.  
Figure~\ref{fig:pix-range} 
plots the number of unique angles $\phi$ versus length $r$ where we note that a line must have approximately 70 pixels or more in order to represent all orientations $\phi=-89,-88,\ldots,90$.  
This exploration of the $(r,\phi)$ space explored 17,820 possible $(r,\theta)$ lines resulting in 13,034 unique $(r_,\phi)_\text{u}$ kernels; this is orders of magnitude larger than other explorations, e.g., the 361 kernels in~\cite{Sun_2015_CVPR}.

\subsection{COCO Blurred Dataset}
\label{sec:dataset}
We use the 2014 COCO dataset~\cite{lin2014microsoft} to create a blurred dataset for training and validating a model for blind estimation of length and angle given only a blurry image.  The 2014 COCO dataset consists of a training dataset with 82,783 images and a validation set of 40,504 images.  The training dataset is used to create a blurred training dataset and the validation dataset is split (with no overlap) to create blurred validation and test datasets.

\begin{figure}[t]
        \centering
        \includegraphics[trim={0 0.7in 0 1in},clip]{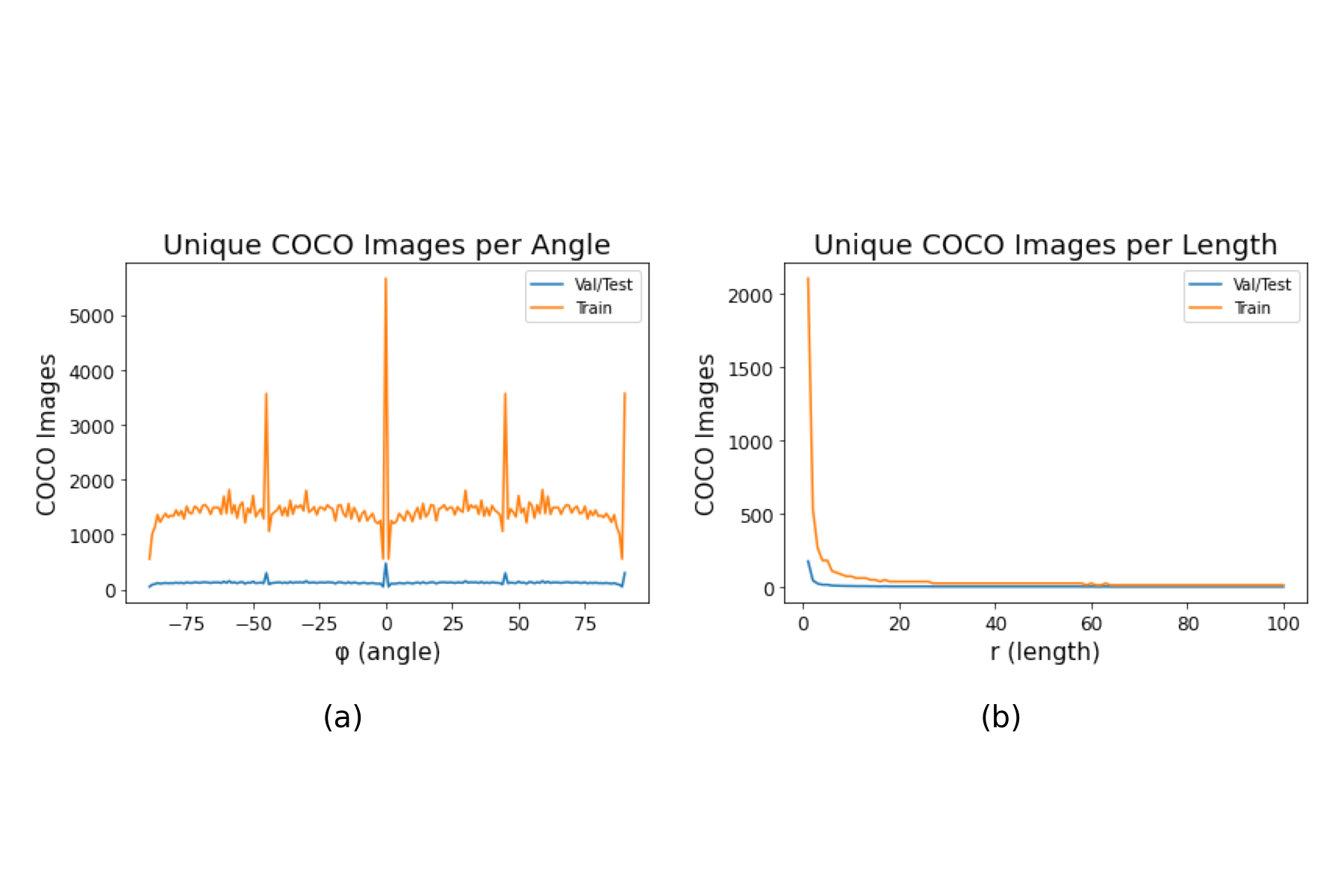}
        \caption{Representation of COCO images versus length and angle. (a) Number of unique COCO images per angle in training and validation/testing datasets. (b) Number of unique COCO images per length in training and validation/testing datasets. } 
        \label{fig:Unique_COCO_Images}
\end{figure} 
   
The blurred dataset is created by defining a blur generator that loops through the labels $(r, \phi)_\text{u}$, creating the corresponding blur kernel using the \texttt{skimage.draw} function (see Sec.~\ref{sec:blur_kernel}), and convolving the blur kernel with an image from the COCO dataset. A random COCO image is chosen (without replacement) for each length $r$ to provide a variety of images in the blurred dataset.
There is a dataset imbalance in lengths and/or angles represented in the blurred dataset due to the uneven representation of angles for shorter lengths (see Fig.~\ref{fig:pix-range}).   We improve the balance by creating a minimum of 175 blurred images per length, implying that more unique COCO images are used for shorter blur lengths.  Each run of the blur generator creates 21,789 blurred images. The creation of the training set used 12 parallel threads of the blur generator, each operating on a subset of the COCO training dataset, creating a total of 261,468 blurred images. The validation and testing sets were each created with one run of the blur generator operating on the validation and testing splits of COCO validation dataset, creating a total of 21,789 blurred images for each. Figure~\ref{fig:Unique_COCO_Images} shows the number of unique COCO images versus length and angle for the training, validation, and testing datasets. We notice a spike in the number of unique images for length $r=1$ and for angle $\phi=0$ due to the non-blurred images that are part of the dataset. In general, more unique images are used for shorter lengths as the blur generator needs to loop through more images to complete the minimum threshold of images per length. We notice also that small length blur kernels create peaks at angles of $\phi=\{-45, 0, 45, 90\}$ since smaller length kernels are limited in the angles they are capable of representing.

\begin{figure}[t]
        \centering
        \includegraphics[trim={0 0.7in 0 1in},clip]{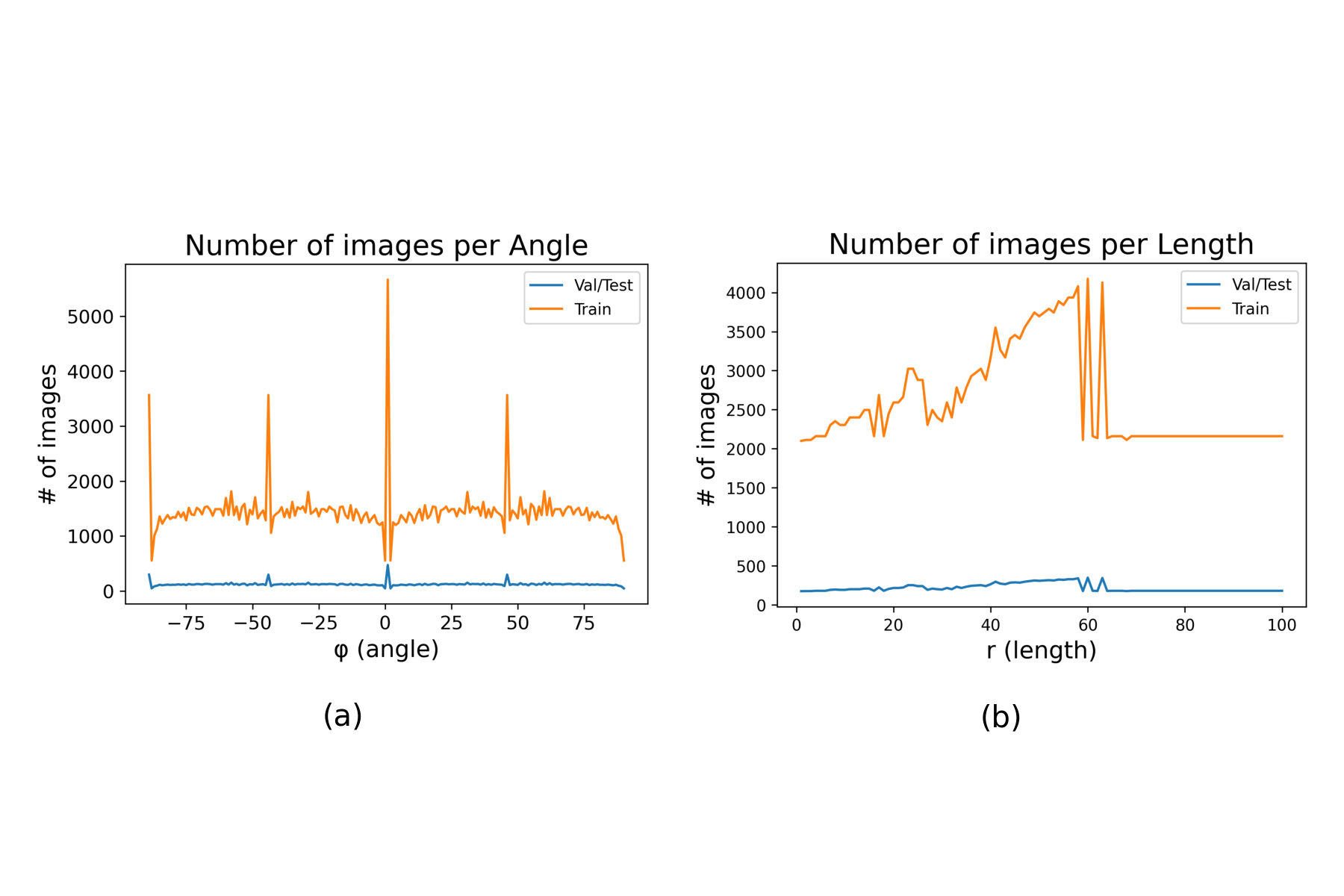}
        \caption{Distribution of blurred images for angle and length. (a) Number of blurred images per angle. (b) Number of blurred images per length.} 
        \label{fig:blurred dataset}
\end{figure}

There is an interdependence between angle and length, resulting in an inability to completely balance both length and angle in the blurred dataset. Figure~\ref{fig:blurred dataset} shows the distribution of labeled blurred images with respect to angle and length. Our choice to generate a minimum of 175 blurred images per length in the blur generator favors a more uniform length distribution [Fig.~\ref{fig:blurred dataset}(b)].  This, however, results in peaks in the angle distribution [Fig.~\ref{fig:blurred dataset}(a)] due to the over-representation of certain angles for smaller lengths (see Fig.~\ref{fig:pix-range}). A choice to favor a more uniform angle distribution, however, would result in smaller representation of shorter blur lengths, creating a more problematic data imbalance.  While there are other choices that could be made regarding data imbalance here, we find that this blurred dataset can be used to train a network to accurately predict both length and angle.

\section{Uniform Blur Prediction}
\label{sec:blur_prediction}
We formulate blur length and angle estimation as a deep learning regression problem using the VGG16~\cite{simonyan2014very} architecture as the backbone of our model. VGG16 was trained from scratch with TensorFlow's default layer initializations to predict both the length and angle parameters for a uniform linear motion blurred image. The last layer of VGG16 was modified to have two output nodes (corresponding to prediction of length and angle) with sigmoid activations.  Since the sigmoid activation outputs values in the range $[0,1]$, length and angle are normalized to be in the range $[0,1]$ to train the model.  The predicted length and angle parameters are re-scaled to their native ranges ($r\in[1,100]$, $\phi\in(-90,90]$) for validation.

Due to the fixed input size of VGG16 ($224\times224$ pixels), we created a TensorFlow data generator to randomly crop the blurred COCO images. We crop instead of resizing the image to maintain accuracy in the blur angle labels since a resize could change the aspect ratio of the image and thus the blur angle. If a blurred image was smaller than the input size for VGG16,  it was skipped and not used in training; in total there were 892 training, 5 validation, and 13 test images that were skipped. This results in less than $0.3\%$ of images skipped for each dataset.

We use the Adam optimizer to with a learning rate of 0.1 and epsilon of 0.1 and the mean squared error (MSE) as the loss function. A batch size of 50 was empirically determined to mitigate convergence issues. The model is trained for 50 epochs, saving the weights for the best model throughout training; training is terminated if the MSE performance has not improved within the previous 5 epochs. Training takes about 25 minutes per epoch and about 12 hours to fully train on an NVIDIA RTX-3090.

We performed multiple trainings of the network for different noise levels, resulting in four different models. The first model is trained with no noise and the other three models are trained with different levels of additive white Gaussian noise with variance $\sigma^2\in\{0.001,0.01,0.1\}$, corresponding to signal-to-noise (SNR) values of $\{30, 20, 10\}$
dB, respectively. Noise is added in our data generator after the blurred image is cropped and normalized to the range $[0,1]$. Each model is tested by predicting blur parameters for the noiseless test set and for three additional test runs in which three noise levels are added to the test set.

\section{Experiments and Results}
\label{sec: Experiment and results}
This section presents results using the $R^2$ coefficient of determination metric for evaluating the regression model. Next, we consider a scenario with additive noise and present results our model's predictions. Finally we compare with other methods of blur kernel estimation~\cite{BD_Krishnan,Levin_2011_ieee,carbajal2021} and evaluate by assessing deblurred images using the error ratio score as introduced by~\cite{levin2009understanding}.

\subsection{Metrics}
We validate the performance of blur estimation using metrics that measure the accuracy of the parameter estimation itself and also the quality of an image deblurred using the estimated kernel.

\subsubsection{Accuracy of Parameter Estimation}
In testing, we measure performance using the coefficient of determination, $R^2$, which measures the goodness of fit between actual known values of variable $y_i$ and estimated values $x_i$:
\begin{equation}
    R^2 = 1-\frac{\sum_{i=1}^n (y_i-x_i)^2}{\sum_{i=1}^n (y_i-\bar{y})^2},
\end{equation}
where $\bar{y}$ is the mean of the known variables $y_i$ and $n$ is the number of samples~\cite{r2}.  The numerator $\sum_{i=1}^n (y_i-x_i)^2$ is the sum of squares of the residual prediction errors and the denominator $\sum_{i=1}^n (y_i-\bar{y})^2$ is proportional to the variance of the known data.  A perfect model will have zero residual errors and thus an $R^2=1$.  A na\"{i}ve model that always predicts the average of the data $\bar{y}$ will have equal numerator and denominator and thus an $R^2=0$.  Models that have predictions worse than the na\"{i}ve model will have an $R^2<0$.

\subsubsection{Quality of Deblurred Image}
In addition to studying the accuracy of blur kernel estimation, we measure the quality of the image deblurred using the estimated blur kernel in a non-blind deblurring method (see Sec.~\ref{sec:results_deblur}).  To measure the quality of the deblurred image, we use the error ratio as presented in~\cite{levin2009understanding}.  The error ratio is motivated by the fact that, even with a perfectly estimated blur kernel, one may not be able to perfectly predict the latent sharp image.  The error ratio $E_{\hat{K}}/E_{\vphantom{\hat{K}}{K}}$ is computed by considering the error $E_{\hat{K}}$ between the true sharp image and the latent sharp image recovered using the estimated blur kernel $\hat{K}$ and the error $E_{\vphantom{\hat{K}}{K}}$ between the true sharp image and the latent sharp image recovered using the true blur kernel $K$.  An error ratio of 1 indicates that the images deblurred using the estimated and true kernel are identical.

We use the sum of squared differences (SSD) as in~\cite{levin2009understanding} to define the error ratio.  Error ratios are generally presented as a cumulative histogram for error ratios binned in the range $[1,4]$.  As noted in~\cite{levin2009understanding}, SSD error ratios above 2 tend to indicate significant perceptually noticeable distortion present in the image deconvolved with the estimated kernel as compared to the image deconvolved with the true kernel.

\subsection{Accuracy of Parameter Estimation}
\subsubsection{Noise-Free Predictions}
\label{sec:noiseless}
\begin{figure}[tp]
        \centering
        \includegraphics[trim={0 0.1in 0 .5in},clip]{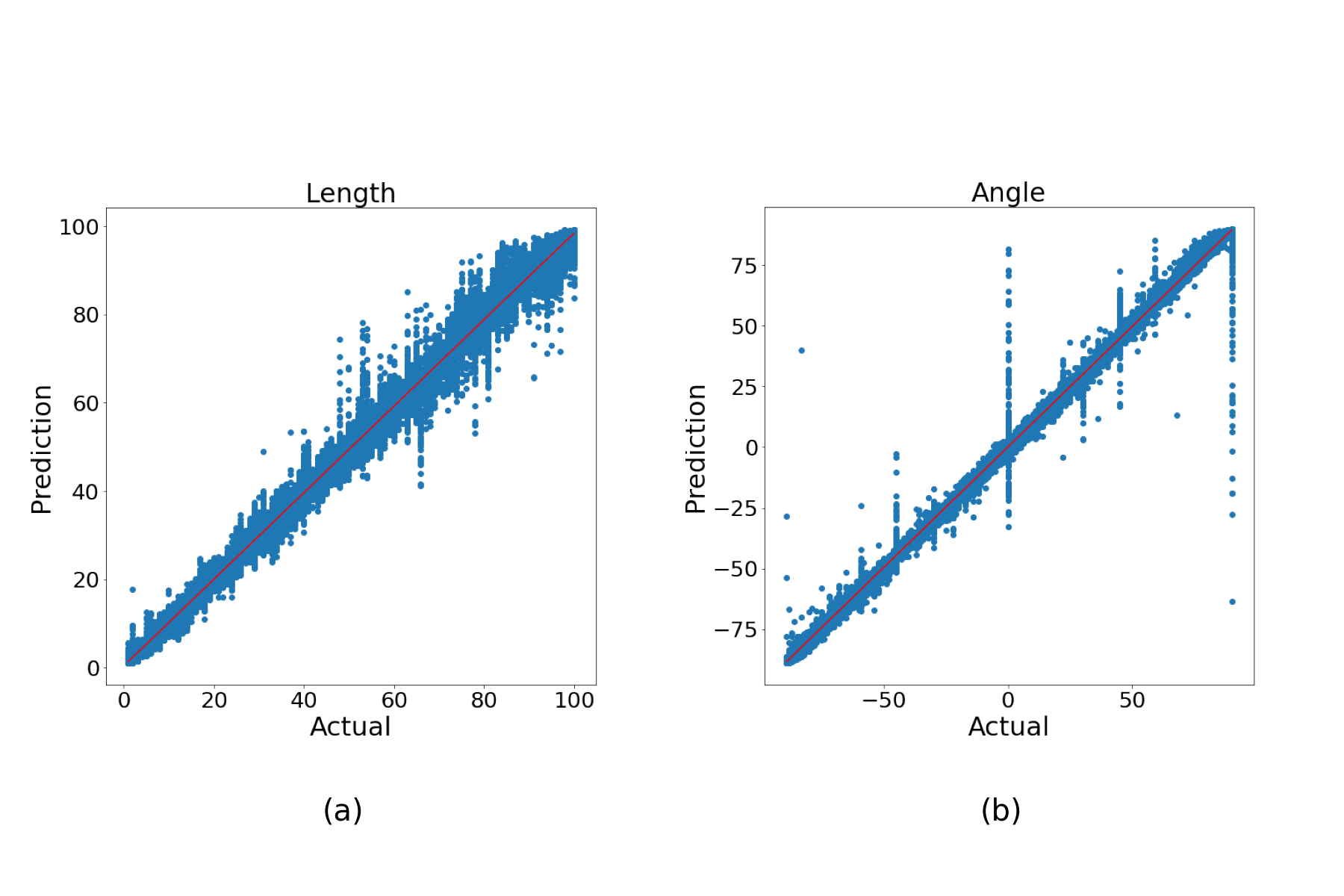}
        \caption{Scatterplots of estimated versus actual values for (a) length and (b) angle.  Individual estimates are represented as blue points and the best fit linear line is plotted in red. (a) Scatterplot of estimated versus actual length, $R^2=0.9869$. (b) Scatterplot of estimated versus actual angle, $R^2=0.9935$.} 
        \label{fig:r2_results}
\end{figure}

A VGG16 based model (Sec.~\ref{sec:blur_prediction}) was first trained on blurred images without any additive noise.  Scatter plots of estimated versus actual length and angle are shown in Fig.~\ref{fig:r2_results} along with the corresponding $R^2$ scores. We find the model to be highly accurate for prediction of both length and angle as demonstrated by the $R^2$ scores of  0.9869 and 0.9935, respectively.  In Fig.~\ref{fig:r2_results}(a) we note a larger spread in estimated values as the length increases.  This is not surprising, as larger blur lengths are expected to be more difficult to accurately estimate.  In Fig.~\ref{fig:r2_results}(b) we note certain angles have a larger spread in estimated values.  This is most notable for $\phi\in\{-45,0,45,90\}$, but can be noted for other angles. This is due to errors in prediction for the smaller length kernels which have a limited set of angles.  It is important to recall, however, that many of these incorrect angle predictions will result in a correct blur kernel.  As an example, an image blurred with kernel parameters $(r,\phi)=(2,0)$ may have an angle prediction of $\hat\phi=27$. However, since blur kernels of length $r=2$ can only be represented by $\phi\in\{-45,0, 45, 90\}$, the blur kernel created with parameters $(r,\phi)=(2,27)$ will generate a blur kernel estimate of $(r,\phi)_\text{u}=(2,0)$ which is correct. 

\subsubsection{Predictions with Additive Noise}
We used the model from Sec.~\ref{sec:noiseless}, trained on noise-free blurred images, and tested it on images with additive white Gaussian noise with variance $\sigma^2\in\{0.001,0.01,0.1\}$, corresponding to SNR values of $\{30,20,10\}$ dB, respectively.  Results for this experiment are shown in the first row of Tables~\ref{table: length} and~\ref{table: angle} for length and angle prediction, respectively.   We note the estimation of the length parameter is more susceptible to noise than angle, resulting in a complete failure of prediction ($R^2$ scores less than 0) for even the smallest level of additive noise, $\sigma^2=0.001$.  We hypothesize that additive noise can alter the intensity distribution along the blur path in a blurry image, creating the appearance of artificially shorter or longer blur paths.  Those blur paths, however, will likely retain more characteristics of their angle for the same level of noise. 

\begin{table}[t!]
\centering
\caption{$R^2$ score for length prediction for training and testing under different levels of additive noise.}
\begin{tabular}{l|llll}
                    & \multicolumn{4}{c}{Testing $\sigma^2$}\\
Training $\sigma^2$ & 0      & 0.001  & 0.01   & 0.1  \\\hline
0                   & 0.9869 & -0.26  & -3.17  & -3.17      \\
0.001               & 0.9607 & 0.9557 & -1.14  & -3.12       \\
0.01                & 0.9527 & 0.9539 & 0.9523 & -2.91        \\
0.1                 & 0.8923 & 0.8932 & 0.8960 & 0.8772    
\end{tabular}
\label{table: length}
\end{table}

\begin{table}[t!]
\centering
\caption{$R^2$ score for angle prediction for training and testing under different levels of additive noise.}
\begin{tabular}{l|llll}
                   & \multicolumn{4}{c}{Testing $\sigma^2$}\\
Training $\sigma^2$ & 0      & 0.001  & 0.01   & 0.1  \\\hline
0                   & 0.9935 & 0.8772 & 0.3935 & 0          \\
0.001               & 0.9758 & 0.9754 & 0.6509 & -0.16       \\
0.01                & 0.9733 & 0.9735 & 0.9682 & 0.0128     \\
0.1                 & 0.8999 & 0.9009 & 0.9010 & 0.8834    
\end{tabular}
\label{table: angle}
\end{table}

We trained three additional models using noisy blurred images with additive white Gaussian noise with variance $\sigma^2\in\{0.001,0.01,0.1\}$ and tested each of those models on noise-free and noisy images, i.e., for $\sigma^2\in\{0,0.001,0.01,0.1\}$.  Results for those three models are in the second through fourth rows of Tables~\ref{table: length} and~\ref{table: angle} for length and angle prediction, respectively.  We again note a higher sensitivity to noise in the prediction of length (Table~\ref{table: length}) than angle (Table~\ref{table: angle}).  We further note that the $R^2$ score decreases by $\sim0.1$ for the model trained on the highest level of noise $\sigma^2=0.1$ and tested on noise-free images, compared to the noise-free model tested on noise-free images, but that the same model is robust to varying levels of noise.  Finally we note that models trained on noisy data appear to be robust to noise levels less than or equal to the noise level on which they are trained.  This implies that a single model trained on a single noise level can yield accurate predictions even for smaller noise levels not seen in training.

Most other methods~\cite{Levin_2011_ieee,Radon_T,Xu_2018_ieee,Li_2018_CVPR} that test under additive noise test up to a Gaussian noise of $\sigma^2=0.01$ to simulate sensor noise, while some~\cite{BD_Krishnan,whyte2012non} test noise up to $\sigma^2=0.02$. In comparison, our model is tested up to $\sigma^2=0.1$ and demonstrates a higher tolerance for noise which can be an advantage when modeling atmospheric turbulence.

\subsection{Quality of Deblurred Images}
\label{sec:results_deblur}
\subsubsection{Deconvolution and Comparison Methods}
We quantify the performance of our blur parameter estimation by using the expected patch log likelihood (EPLL) method~\cite{zoran2011learning} to deconvolve the images using our predicted motion blur kernel and the ground truth blur kernel. We used the Python implementation available at~\cite{EPLL_Github}, replacing the Gaussian kernel with a linear motion blur kernel.  EPLL only accepts square, odd-sized blur kernels, necessitating a padding of kernels to be square and odd-sized.  We symmetrically zero pad the shortest side of the kernel to keep the kernel centered and pad to match the size of the longest side. If the longest side is even, however, this results in an even-sized square kernel. We use a similar approach to~\cite{evenKernel}, creating four odd-sized kernels each zero padded with the line asymmetrically offset toward a different corner. The blurred image is deconvolved with each of the four kernels and the average of the four deconvolved images is considered the resulting deblurred image. 

Additionally, we compare performance with the blur kernel estimation methods from Levin et al.~\cite{Levin_2011_ieee}, Krishnan et al.~\cite{BD_Krishnan}, and Carbajal et al.~\cite{carbajal2021}  These works also include methods for estimating the deblurred image.  For fairness of comparisons, we use only the blur kernel estimate from~\cite{Levin_2011_ieee,BD_Krishnan,carbajal2021} and use those blur kernel estimates in the same EPLL~\cite{zoran2011learning} deblurring framework as described above.  For the method in~\cite{Levin_2011_ieee}, we used the Matlab implementation provided at~\cite{Levin_Github}, using the \verb+deconv_diagfe_filt_sps+ function with a kernel prediction size of $(101,101)$ to allow prediction of the largest kernel size in our blurred dataset.  Since the method in~\cite{Levin_2011_ieee} estimates a blur kernel for each of the three channels in a color image, we applied EPLL to each channel using the kernel estimated for that channel. For the method in~\cite{BD_Krishnan}, we used the Matlab implementation provided at~\cite{Krishnan_Github}, again with kernel size of $(101,101)$.  For the method in~\cite{carbajal2021}, we used the python implementation provided at~\cite{carbajal_github} with the fixed kernel size of $(33,33)$; we note that this will put this method at a disadvantage in predicting longer blur lengths.  Additionally, we estimated a single blur kernel for the uniformly blurred images by using the image-averaged mixing coefficients in the superposition of the kernel bases; this allows us to use the same EPLL deblurring framework for comparison to the other methods. 

\subsubsection{Reduced Test Set}
Due to the computational complexity of the EPLL deconvolution~\cite{zoran2011learning}, as well as the kernel estimation methods in~\cite{BD_Krishnan,Levin_2011_ieee,carbajal2021}, we generated reduced-size test datasets for these experiments. From the test dataset (Sec.~\ref{sec:dataset}), two subsets are created to span length and angle, respectively. The first subset, subsequently referred to as Length 5 (L5), uses $5$ randomly selected blurred images for each length, totaling $500$ images. The second subset, subsequently referred to as Angle 3 (A3) uses $3$ randomly selected blurred images for each angle, totaling $540$ images. All images in these reduced test sets are noise free.  

\subsubsection{Error Ratio Comparisons}
\begin{figure}[t]
        \centering
        \includegraphics[trim={0 0.75in 0 0.75in},clip]{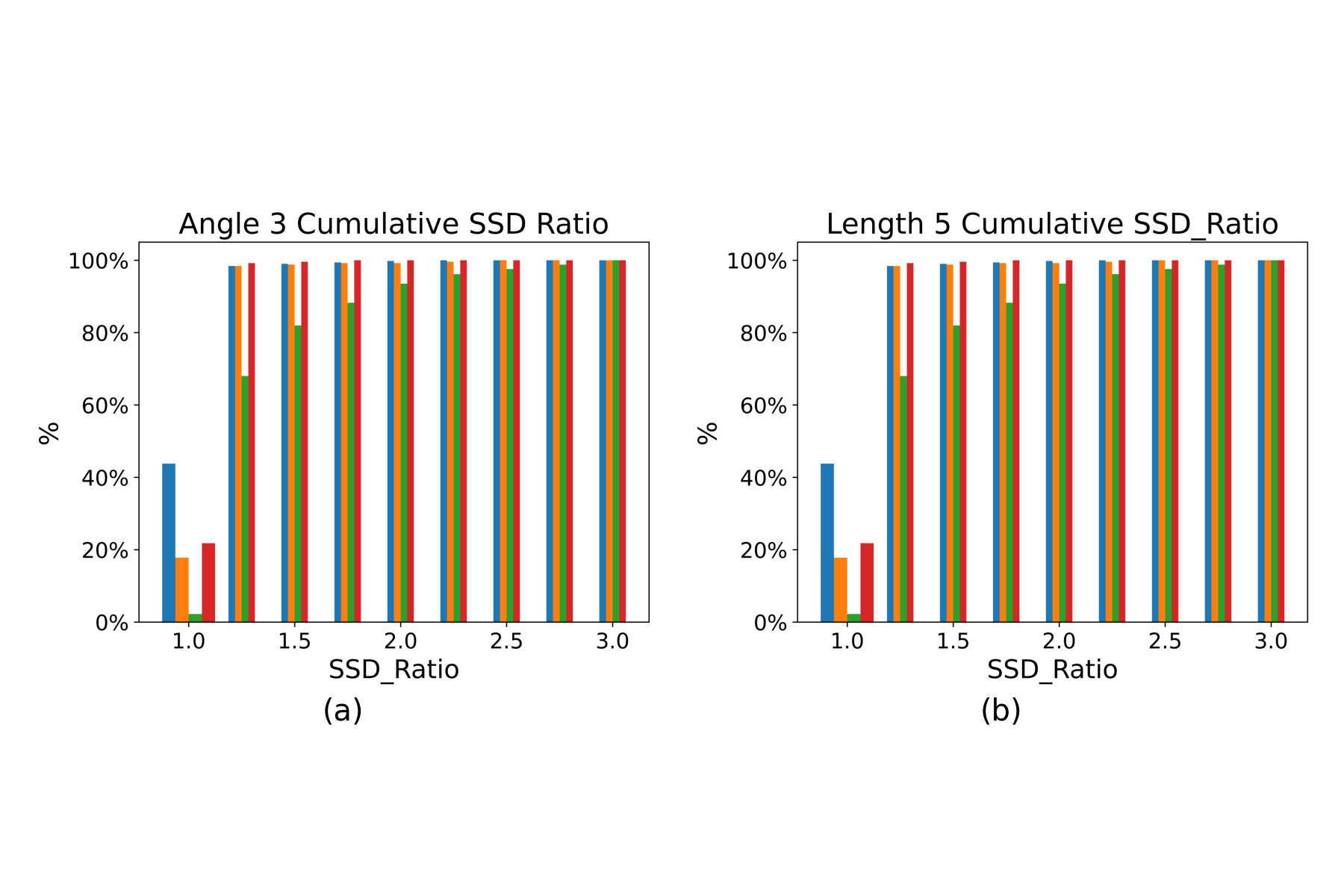}
        \caption{Error ratios for the (a) A3 sub-dataset and (b) L5 sub-dataset.  The bin at 1.0 includes error ratios $\le1.0$ and the bin at 3.0 includes error ratios $\ge3.0$. The results are presented for the proposed method (\textcolor{mBlue}{\solidXXLrule[3.5mm]}), Krishnan et al.~\cite{BD_Krishnan} (\textcolor{mOrange}{\solidXXLrule[3.5mm]}), Levin et al.~\cite{Levin_2011_ieee} (\textcolor{mGreen}{\solidXXLrule[3.5mm]}), and Carbajal et al.~\cite{carbajal2021} (\textcolor{mRed}{\solidXXLrule[3.5mm]}).} 
        \label{fig:SSD}
\end{figure}

We calculated SSD error ratios for our proposed blur kernel estimation as well as those from Levin et al.~\cite{Levin_2011_ieee}, Krishnan et al.~\cite{BD_Krishnan}, and Carbajal et al.~\cite{carbajal2021} as seen in Fig.~\ref{fig:SSD}. Recall that an error ratio of 1 indicates that deblurring with the estimated kernel results in an identical image to that deblurred with the ground truth kernel.  An error ratio of 2 or higher is considered unacceptable as it was shown in~\cite{levin2009understanding} that such SSD error ratios indicate the presence of significant perceptually noticeable distortions in the latent image estimated using the estimated kernel.  An error ratio $<1$ indicates that the image deblurred with the estimated kernel achieves a better match to the true sharp image than deblurring with the true kernel.

Our proposed method has the highest cumulative histogram values for the error ratio at 1 for both A3 and L5 datasets and yields an error ratio of 1.25 or less for most images in the A3 and L5 test sets. This means that our model is able to predict more accurate kernels compared to the other methods. Since our method creates the kernel using linear parameters, we are less prone to additive noise in the kernel. Levin et al.'s~\cite{Levin_2011_ieee} kernel prediction has noise added in the kernel since many of the pixels that are supposed to be zero are instead small numbers close to zero. This noise can be seen to affect its results in kernel prediction. Krishnan et al.~\cite{BD_Krishnan} thresholds the small elements of the kernel to zero which increases robustness to noise.  Carbajal et al.~\cite{carbajal2021} has competitive performance for error ratios $\ge$1.25; the method is at a disadvantage at longer blur lengths due to its limitation to $33\times33$ kernels as noted above which may contribute to the diminished performance at the lowest error ratio.

\begin{figure}[tp]
    \centering
    \includegraphics[]{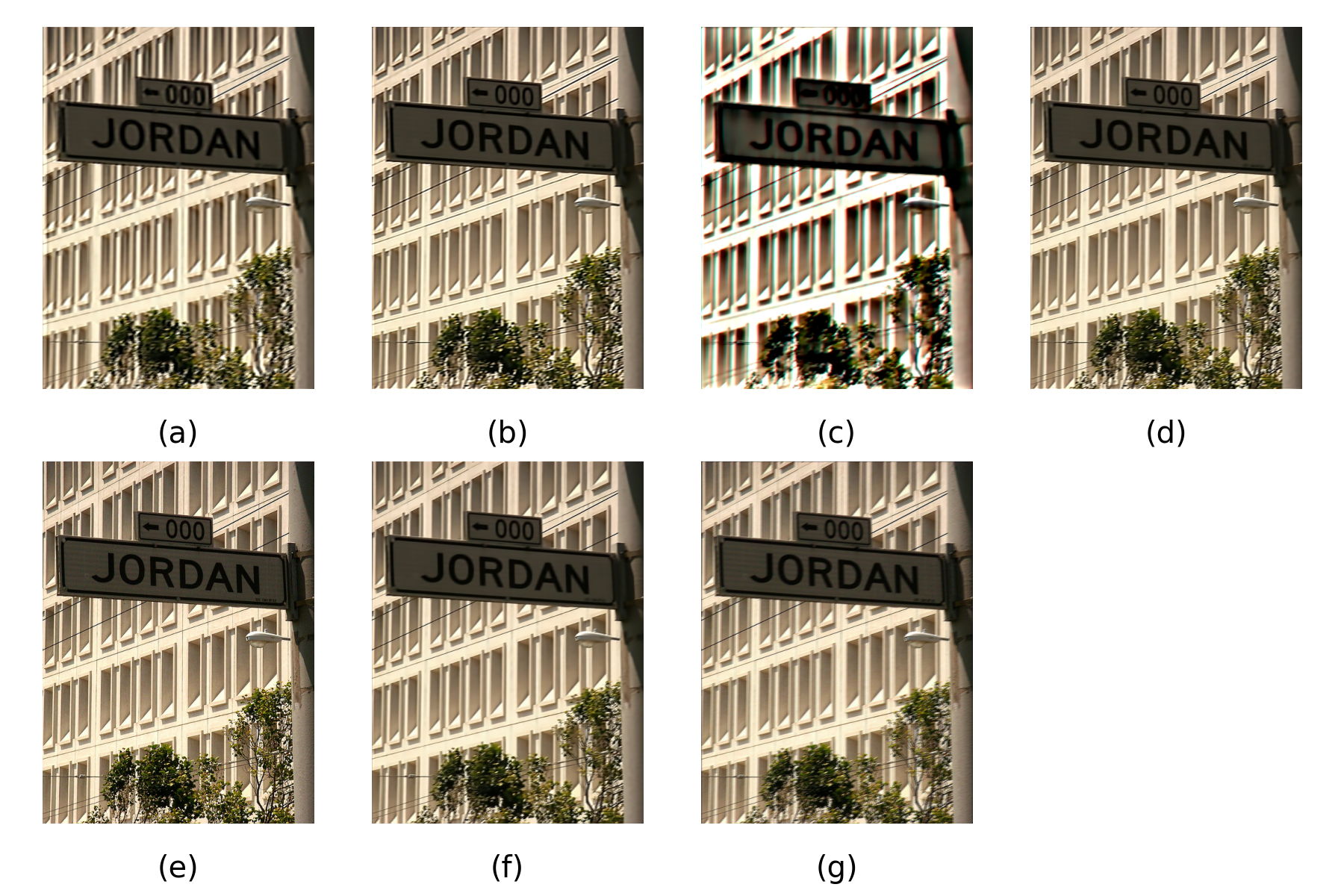}
    \caption{Deblurred images for image blurred with a kernel of length 5 and angle 17.  All images are deblurred with the EPLL method~\cite{zoran2011learning}.  (a) Image deblurred using kernel estimated with our proposed method, error ratio 1.2858. (b) Image deblurred using kernel estimated with the method in Krishan et al.~\cite{BD_Krishnan}, error ratio 1.1320. (c) Image deblurred using kernel estimated with the method in Levin et al.~\cite{Levin_2011_ieee}, error ratio 1.5607. (d) Image deblurred using kernel estimated with the method in Carbajal et al.~\cite{carbajal2021}, error ratio 1.0396. (e) Original sharp image.  (f) Image deblurred with the ground truth kernel. (g) Blurred image.}
    \label{fig:deblurred 1}
\end{figure}

\begin{figure}[t]
    \centering
    \includegraphics[]{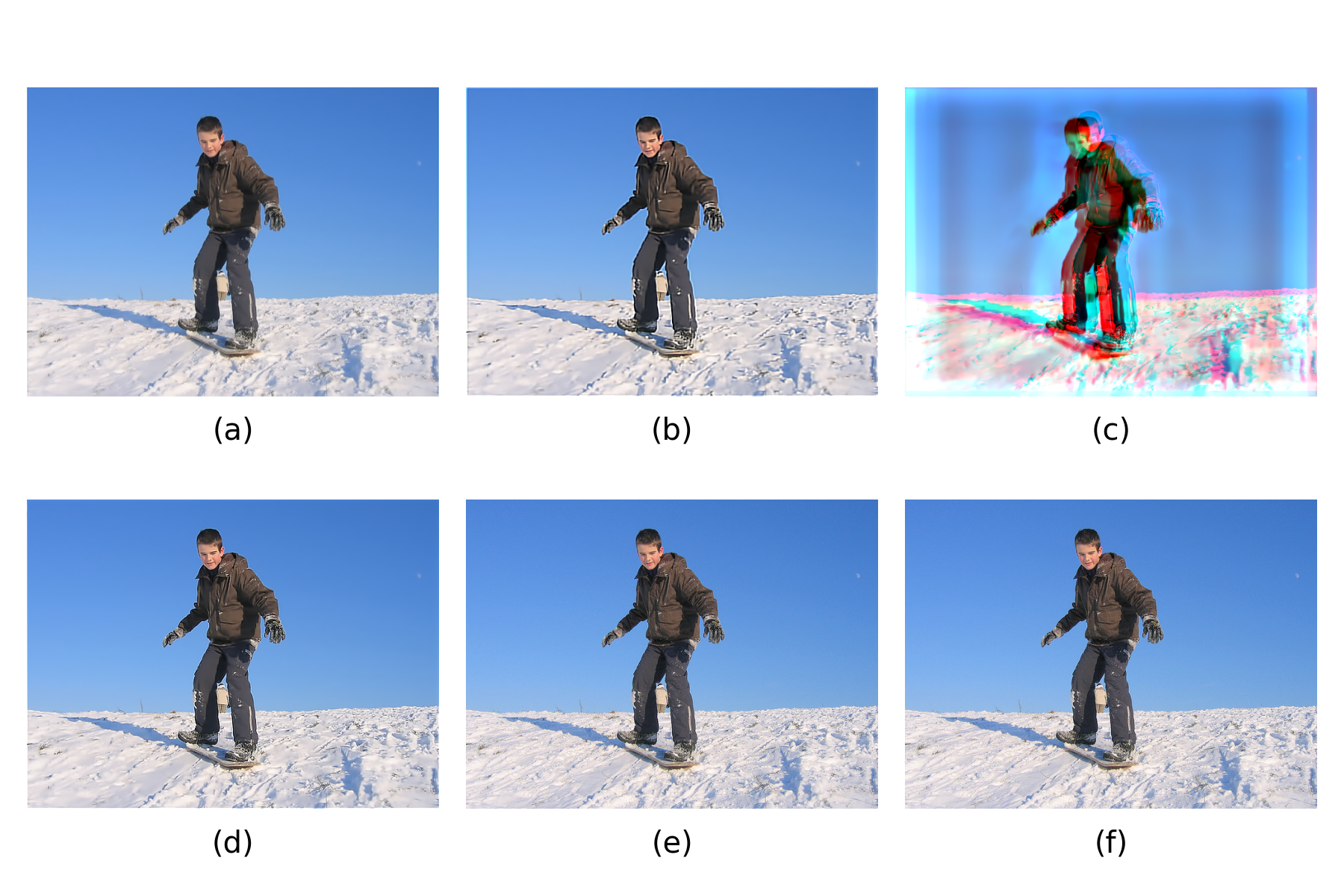}
    \caption{Deblurred images for sharp input image.  All images are deblurred with the EPLL method~\cite{zoran2011learning}.  (a) Image deblurred using kernel estimated with our proposed method, error ratio 1.5813. (b) Image deblurred using kernel estimated with the method in Krishan et al.~\cite{BD_Krishnan}, error ratio 1.7345. (c) Image deblurred using kernel estimated with the method in Levin et al.~\cite{Levin_2011_ieee}, error ratio 4.4847. (d) Image deblurred using kernel estimated with the method in Carbajal et al.~\cite{carbajal2021}, error ratio 1.3388. (e) Original sharp image.  (f) Image deblurred with the ground truth kernel.}
    \label{fig:deblurred 2}
\end{figure}

Figures~\ref{fig:deblurred 1} and~\ref{fig:deblurred 2} show two qualitative examples of images deblurred using the kernels estimated with our proposed method and the methods in~\cite{BD_Krishnan,Levin_2011_ieee,carbajal2021}. The example in Fig.~\ref{fig:deblurred 1} uses a blurred image as an input and the example in Fig.~\ref{fig:deblurred 2} has a sharp image as an input for demonstration of the methods in the absence of blur.  We note similar qualitative results between the deblurred images using our estimated blur kernel and the blur kernel estimated by the method in~\cite{BD_Krishnan,carbajal2021}, and that both methods yield similar results to the image deblurred with the truth kernel.  We do note, however, some ringing in the images deblurred using the kernel from~\cite{BD_Krishnan}; this is particularly noticeable in the areas around the upper power line in the image in Fig.~\ref{fig:deblurred 1}(b) and the darker snow shadows in Fig.~\ref{fig:deblurred 2}(b).  The blur kernel estimated by the method in~\cite{Levin_2011_ieee} introduces significant artifacts, especially apparent in Fig.~\ref{fig:deblurred 2}(c) but also apparent in Fig.~\ref{fig:deblurred 1}(c) as dark regions in the sign. The blur kernel estimated by~\cite{carbajal2021} yields very good qualitative results, perhaps indicating the advantage of a kernel that is not constrained to linear motion blur when considering very large image regions.  Overall, the proposed method of kernel estimation appears to yield deblurred images close to that which could be achieved with the ground truth kernel which validates this approach as the foundation for future work in estimation of motion-blur parameters in spatially-varying blur.  

\section{Conclusions}
\label{sec:conclusions}
In this paper, we have studied in detail the limitation in representation of linear blur kernels, particularly for shorter blur lengths.  We find an interdependence between length and angle in representing blur kernels, meaning that developing a dataset that is balanced in both length and angle is not possible.  Furthermore, while much existing research in linear blur prediction has implicitly assumed square odd-sized kernels, we relax this assumption and allow for non-square even-sized kernels. A blurred dataset was created from the 2014 COCO dataset using a suite of blur kernels for length $r\in[1,100]$ and angle $\phi\in(-90,90]$, providing a more comprehensive variety of blur kernels than previously studied.  

This thorough definition of linear motion blur and the development of a blurred dataset allowed us to train a regressive deep learning model instead of a classification model as other deep learning methods implement. With regression we can estimate more accurate blur kernel parameters which are not limited by the model needing \textit{a priori} information about the kernel size. We demonstrate excellent performance in estimation of blur kernel parameters with a coefficient of determination $R^2\ge0.89$ for both length and angle prediction.  The robustness of the estimation to additive noise was studied for a wider range of noise than previously considered, $\sigma^2\in\{0.001,0.01,0.1\}$, corresponding to SNR values of $\{30,20,10\}$ dB.  The regression CNN was found to be very robust to noise, with a $10\%$ drop in the $R^2$ metric for a $10\%$ Gaussian noise (which we note is an order of magnitude larger than the noise level of 1\% commonly studied).  We further note that the model trained on 10\% Gaussian noise was robust to noise levels less than 10\%, indicating that single model can serve across a wide range of noise scenarios.  Using the estimated blur kernels in a non-blind deblurring method, we find sum of SSD error ratios of 1.25 or less for most test images, significantly outperforming the MAP-based comparison methods and competitive with a recent deep-learning based method.  

In future work, we will use this exploration of linear motion blur kernels as a baseline and foundation for spatially-varying blurs.  We will expand this to a patch level approach to decompose a spatially-varying blur image into a superposition of locally uniform blur patches. Extending the approach to non-uniform blur will provide a better means to estimate the spatially varying blur resulting from atmospheric turbulence in images.  Additionally, it would be interesting to directly compare our approach to other deep learning approaches which either directly estimate length and angle or which assume locally linear motion blur kernels, e.g.,~\cite{Sun_2015_CVPR, nasonov2022, yan_2016_ieee, Gong_2017_CVPR}.

\subsection*{Disclosures}
The authors have no conflicts of interest to disclose.

\subsection* {Code, Data, and Materials Availability} 
The code necessary to implement the proposed method of kernel estimation as well as to generate the blurred dataset from the publicly available COCO dataset is available GitHub repository at \url{https://github.com/DuckDuckPig/Regression_Blur}.

\subsection* {Acknowledgments}
The authors gratefully acknowledge Office of Naval Research grant N00014-21-1-2430 which supported this work.


\bibliography{references}   
\bibliographystyle{ieeetr}   

\end{document}